\documentclass[useAMS,usenatbib]{mn2e}
\usepackage{graphics,epsfig,lscape,txfonts,color}
\usepackage[english]{babel}
\usepackage{caption}

\newcommand{\mr}{\mathrm}

\def\ie{i.\,e.}                                      
\def\eg{e.\,g.}                                      

\def\xmm{\textit{XMM-Newton}}

\def\gx339{GX\,339-4}
\def\h1743{H\,1743-322}

\def\grs{GRS\,1915+105}

\title[Distinct soft and hard band PDS in GRS 1915+105]{Detection of distinct power spectra in soft and hard X-ray bands in the hard state of \grs\thanks{Based on observations obtained with \xmm, an ESA science mission with instruments and contributions directly funded by ESA Member States and NASA.}}
\author[H. Stiele, W. Yu]{H. Stiele$^{1}$\thanks{E-mail:
hstiele@shao.ac.cn}, W. Yu$^{1}$ \\
$^{1}$Shanghai Astronomical Observatory and Center for Galaxy and Cosmology, 80 Nandan Road, Shanghai, 200030, China}
\begin{document}

\date{2014 March 3}

\pagerange{\pageref{firstpage}--\pageref{lastpage}} \pubyear{2014}

\maketitle

\label{firstpage}

\begin{abstract}
The well-known black hole X-ray binary \grs\ is a unique source in the sense that it cannot be classified within the standard picture of black hole binary states.\\ 
In this work we study archival \xmm\ observations taken between 2003 and 2004 of the $\chi$ variability class of \grs, which corresponds to the hard state in the standard black hole X-ray binary state classification. The crucial point of our study is that by using \xmm\ data we can access the variability below 3 keV, an energy range that is not covered with RXTE. We focus on the study of the power spectral shape in the soft and hard X-ray band, in light of our work done with Swift on MAXI J1659-152.\\ 
In the hard band (above 2.5 keV) power density spectra consist of band-limited noise and quasi-periodic oscillations, corresponding to the power spectral shape seen in the hard or intermediate state, while in the soft band the averaged power density spectrum is consistent with a power-law noise, corresponding to the power spectral shape usually seen in the soft state. The coexisting of two different power spectral shapes in the soft and hard band, where the soft band power spectrum is dominated by a power-law noise, is consistent with MAXI J1659-152, and confirms the energy dependence of power spectral states. Our additional spectral analysis shows that the disc component does contribute to the soft band flux.\\
These findings support that the observed black hole power spectral state depends on which spectral component we are looking at, which implies that power spectral analysis is probably a more sensitive method than spectral modeling to trace the emergence of the disc component in the hard or intermediate state. 
\end{abstract}


\section{Introduction}

The known population of low-mass black hole X-ray binaries mainly consists of transient sources, that can be studied only during outburst, as they are too faint to detect their variability reliably with present X-ray instruments during quiescence  \citep[see e.\,g.][]{1998ASPC..137..506G}. The outbursts begin and end in the so-called low-hard state (LHS) and in between there is normally a transition to the high-soft state (HSS). All these states show characteristic timing and spectral properties. In the LHS the rms variability is larger than ten per cent and the power spectrum shows one or more band-limited noise (BLN) components and sometimes a specific timing feature named type-C quasi-periodic oscillations (QPOs) can be observed \citep{2011BASI...39..409B}.  The power spectrum of the HSS is well-described by a power-law noise (PLN) component, sometimes with a break around 10 Hz, and the rms variability is at a few percent \citep{2001ApJS..132..377H}.

In a recent study \citep{2013ApJ...770..135Y}, we showed an energy dependence of the power spectra in the black hole candidate MAXI J1659-152. The source was unique in several respects. The spectral evolution was slow and the source was at a high latitude, allowing us to study the emergence of the soft disc component and how the soft component came in just before the transition to the soft state. We investigated energy and power density spectra of \textit{Swift}/XRT (0.3 -- 2 keV) and RXTE/PCA (2 -- 60 keV) observations that covered the outburst rise from the LHS to the HSS. During the LHS the power density spectra in the 0.01 to 20 Hz range can be well described by BLN and QPOs in both, the soft and the hard, X-ray bands. With the onset of the hard intermediate state, which coincided with a disc fraction exceeding $\sim$30\% in the 0.3 -- 2.0 keV range\footnote{This fraction has to be taken with caution as it is obtained from fitting a disc blackbody plus power-law model to the data, without taken into account that the power-law model should have a low energy cutoff to mimic Comptonization.}, this changed dramatically. While BLN and QPOs are still present above 2 keV, below 2 keV the power spectra are now dominated by PLN, as commonly observed in the HSS.  This suggests that the photons responsible for the BLN and the QPO origin from the innermost hot flow subjected to Comptonization, while the photons responsible for the PLN can be related to the optically thick disc. Furthermore, we tried to constrain cut-off energies for the PLN and BLN plus QPO components, investigating contributions of each component in the 2 -- 4 keV XRT band and at PCA energies below 5 keV. Based on these investigations we could constrain the cut-offs to occur in the 2.8 -- 3.5 keV range.

The well known black hole low mass X-ray binary \grs\ \citep[for a review see][]{2004ARA&A..42..317F} was initially discovered by the \emph{WATCH} instrument on-board \emph{GRANAT} in 1992 \citep{1992IAUC.5590....2C}. Since then, \grs\ has been observed densely at different wavelength ranges. A systematic monitoring in the X-rays revealed a rich pattern of variability on all time scales. \citet{2000A&A...355..271B} identified 12 classes of variability and showed that, though complex, the behaviour of \grs\ can be understood as transitions between three basic spectral states A, B, C. Despite its many distinct accretion states, \grs\ appears similar to other black hole binaries \citep[][and references therein]{2003A&A...412..229R,2010MNRAS.409..763V}. The closest analogue to the conventional canonical ``low hard" state in other X-ray binaries is the $\chi$ variability class, that is found exclusively in the C state \citep{2013ApJ...778..136P}. In this state low frequency QPOs are present \citep{1997ApJ...477L..41C}, which energy spectrum consists with that of the hard component \citep{1999ApJ...513L..37M}. Correlations of the centroid frequency with the power-law index \citep{2003A&A...397..729V} and with the inner-disc radius \citep{2002A&A...386..271R,2002A&A...387..487R} have been conducted and it was shown that the centroid frequency correlates positively with the flux of the disc component \citep{1999ApJ...527..321M}.
For completeness, we would like to mention that with IGR J17091-3624, a second source is known, that shows variability similar to the variability classes observed in \grs\ \citep{2011ApJ...742L..17A,2012MNRAS.422L..91W,2012MNRAS.422.3130C,2013MNRAS.436.2334P,2013ApJ...778...46P}.
 
In this paper we make use of archival \xmm\ observations to study the properties of power density spectra in different energy bands. Specifically, we intend to study the power spectra of the soft disc component in the energy range below $\sim$2 keV and to check if the soft band variability in the $\chi$ class is indeed not only different from that in the hard band 
but also shows a power-law noise component similar to that in the soft state.

\section[]{Observations and data analysis}
\subsection{\xmm}
\label{Sec:obs}
\xmm\ observed \grs\ several times in 2003, 2004 \citep{2006A&A...448..677M}, and 2007. All but one of these observations are taken with the pn detector in burst mode. This mode was chosen because of the high source flux of \grs. The only observation which has been taken with EPIC/pn in Timing mode (2004 April 17) suffers from frequent telemetry drop-outs and can hence not be used in our timing study \citep[see also][]{2006A&A...448..677M}.

For our study we selected five observations of \grs\ being in the $\chi$ variability class. Details on these observations can be found in Table~\ref{Tab:Obs}. We used the standard SAS (version 13.0.0) tools to filter and extract pn event files, paying particular attention to extract the list of photons not randomized in time. For our timing study we selected the longest, continuous exposure available in each observation (see table~\ref{Tab:Obs}), \ie\ the longest available standard good time interval. As all observations are taken in burst mode, we selected photons from a 15 column wide strip in RAWX centered on the column with the highest count rate, and we impose RAWY $<$ 150 to avoid direct illumination by the source. We selected single and double events (PATTERN$<=$4). We made use of the SAS task \texttt{epatplot} to investigate whether the observations are affected by pile-up. As there is a clear deviation from the theoretical predictions at energies below 1.5 keV, we focused our investigations on energies above 1.5 keV, where the observed pattern distributions follow the theoretical predictions. Applying this selection we only exclude 3 -- 4 \% of the source photons, as \grs\ is highly absorbed at energies below 1.5 keV \citep[see][]{2006A&A...448..677M}. All values given in this paper are 1 $\sigma$ values.
 
\begin{table}
\caption{Details of \xmm\ observations}
\begin{center}
\begin{tabular}{rrlrr}
\hline\noalign{\smallskip}
 \multicolumn{1}{c}{No.} & \multicolumn{1}{c}{Obs.~id.} & \multicolumn{1}{c}{Date}  &  \multicolumn{1}{c}{Net Exp. [ks]}  &  \multicolumn{1}{c}{Exp.$^{\dagger}$ [ks]}  \\
 \hline\noalign{\smallskip}
A & 0112990101 & 2003 March 29 & 7.60 & 7.50 \\
B & 0112920701 & 2003 April 10 & 6.09 & 6.09 \\
C & 0112920801 & 2003 April 16 & 1.46 & 1.37 \\
D & 0144090201 & 2004 April 21 & 20.94 & 19.00 \\
E & 0112921201 & 2004 May 3 & 18.75 & 18.70 \\
\hline\noalign{\smallskip} 
\end{tabular} 
\end{center}
Notes: $^{\dagger}$: longest continuous exposure available
\label{Tab:Obs}
\end{table}

\begin{table}
\caption{Details of RXTE observations}
\begin{center}
\begin{tabular}{rrlrr}
\hline\noalign{\smallskip}
 \multicolumn{1}{c}{No.} & \multicolumn{1}{c}{Obs.~id.} & \multicolumn{1}{c}{Date}  &  \multicolumn{1}{c}{Exp. [ks]}  &  \multicolumn{1}{c}{XMM obs.$^{\dagger}$}  \\
 \hline\noalign{\smallskip}
1 & 80127-02-03-00 & 2003 April 10 & 12.55 & B \\
2 & 70702-01-50-00 & 2003 April 16 & 3.37 & C \\
3 & 90108-01-06-00 & 2004 May 3 & 1.29& E \\
\hline\noalign{\smallskip} 
\end{tabular} 
\end{center}
Notes: $^{\dagger}$: corresponding \xmm\ observation
\label{Tab:ObsRxte}
\end{table}

\begin{figure*}
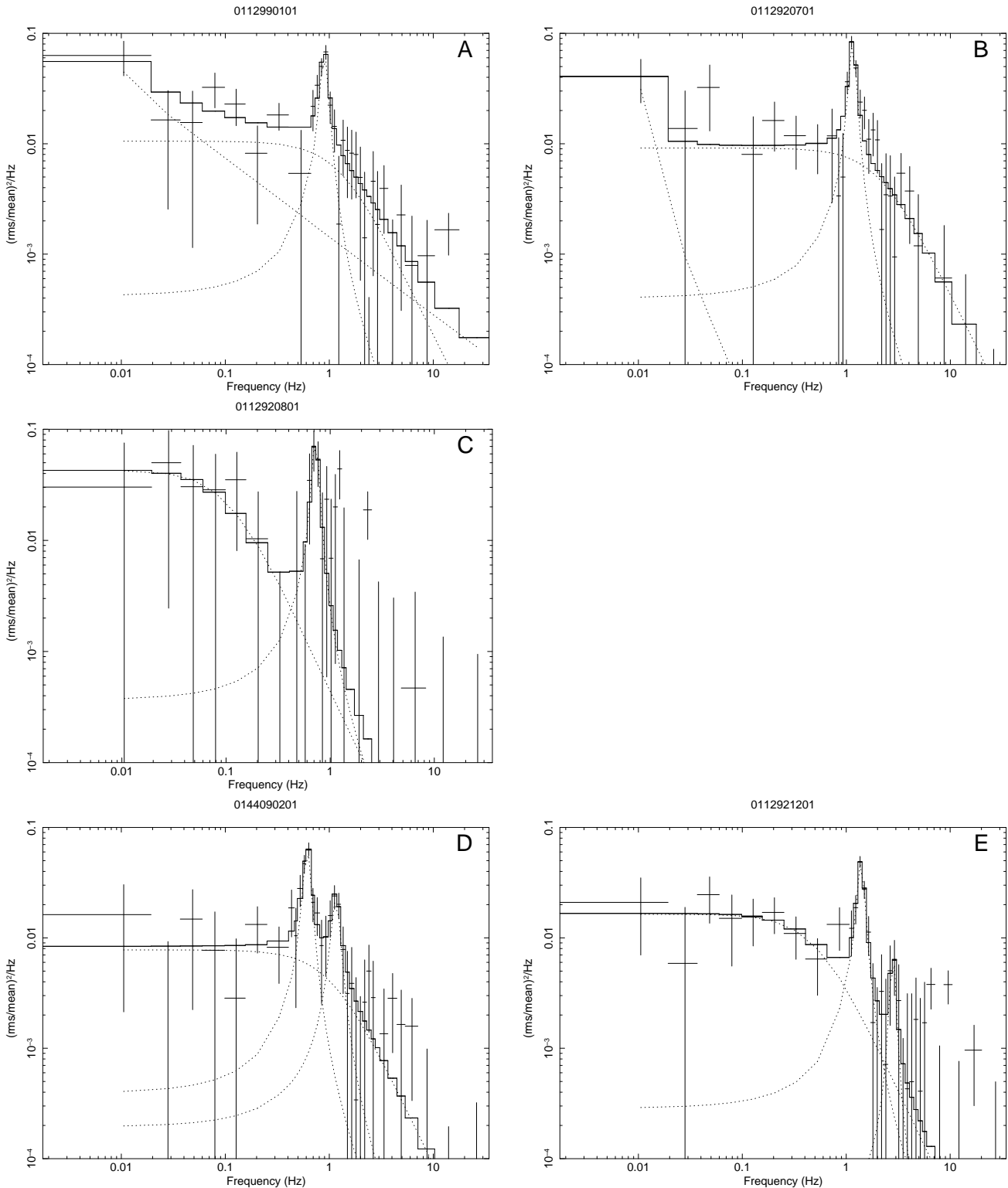

\resizebox{\hsize}{!}{\includegraphics[clip,angle=-90]{plot_PDS_45_8_2003_1.ps}\hskip0.2cm\includegraphics[clip,angle=-90]{plot_PDS_45_8_2003_2.ps}}
\resizebox{\hsize}{!}{\includegraphics[clip,angle=-90]{plot_PDS_45_8_2003_3.ps}\hskip24.4cm}\\
\resizebox{\hsize}{!}{\includegraphics[clip,angle=-90]{plot_PDS_45_8_2004_2.ps}\hskip0.2cm\includegraphics[clip,angle=-90]{plot_PDS_45_8_2004_3.ps}}
\caption{Power density spectra of all five \xmm\ observations in the 4.5 -- 8 keV band. The best fit model is indicated by a solid line and the individual components are given as dashed lines.}
\label{Fig:PDS}
\end{figure*}

\begin{figure*}
\resizebox{\hsize}{!}{\includegraphics[clip,angle=0]{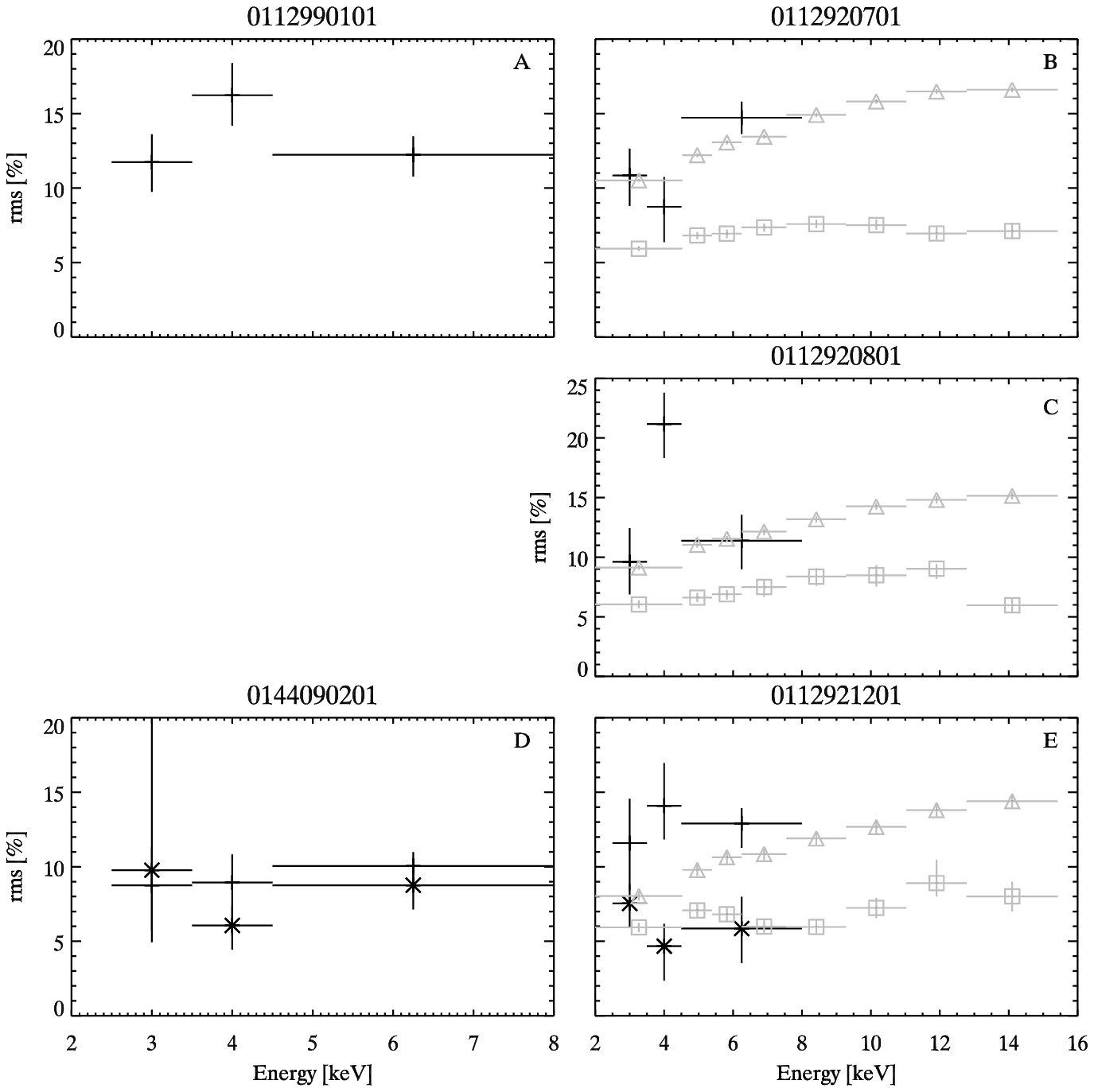}}
\caption{Shown are rms spectra of the QPO of the five \xmm\ $\chi$ variability class observations investigated in this study (frequency range: $1.8\times10^{-3}$ -- 34 Hz). For both observations taken in 2004, the rms spectrum of the main QPO, as well as its upper harmonic (indicated by an "X") is given. For comparison the rms spectra of the main QPO and its upper harmonic derived from RXTE data are indicated by gray triangles and squares.}
\label{Fig:rmsspec}
\end{figure*}

\subsubsection{Timing analysis}
We extracted power density spectra (PDS) for each observation in four energy bands: 1.5 -- 2.5 keV, 2.5 -- 3.5 keV, 3.5 -- 4.5 keV, and 4.5 -- 8.0 keV. These four energy bands contain about 15\%, 20\%, 20\%, 35 -- 40\% of the source photons, respectively. We investigated the noise level at frequencies above 30~Hz and found that it is consistent with 2, as expected for Poissonian noise \citep{1995ApJ...449..930Z}. We subtracted the contribution due to Poissonian noise, normalised the PDS according to \citet{1983ApJ...272..256L} and converted to square fractional rms \citep{1990A&A...227L..33B}. As in \citet{2013ApJ...770..135Y} the PDS were fitted with models composed of a power-law noise, zero-centered Lorentzians for BLN components, and Lorentzians for QPOs.  

\subsubsection{Spectral investigation} 
\label{SubSec:XMMSpec}
We also extracted energy spectra from all \xmm\ observations, following the procedure to extract spectra from \xmm\ burst mode data outlined in \citet{2006A&A...453..173K}. As for the timing study we selected source photons from a 15 column wide strip in RAWX centered on the column with the highest count rate, including single and double events. Of course, no energy selection has been applied here, and we used RAWY $<$ 140 as suggested by \citet{2006A&A...453..173K}. Background spectra have been extracted from columns 3 to 5. 

Energy spectra obtained form \xmm\ EPIC-pn fast-readout modes are known to be affected by gain shift due to Charge-transfer inefficiency (CTI) which leads to an apparent shift of the instrumental edges visible at low energies. This shift can be corrected by applying the SAS task \textsc{epfast} to the data. However,  \textsc{epfast} is likely unsuited to do CTI corrections at higher energies at present, as it applies an energy-independent correction, which leads to an over-correction at higher energies. This leads to a striking difference between the RXTE/PCA and the \textsc{epfast}-corrected \xmm/pn spectra. In contrast, the unmodified \xmm\ spectrum is quite similar to the RXTE/PCA spectrum. This deviation of the spectral shape above $\sim$6 keV due to the application of \textsc{epfast} has been already reported for simultaneous \xmm\ and \textit{BeppoSAX} data \citep{2012MNRAS.422.2510W}. 

\subsection{RXTE}
For three of the \xmm\ observations (Obs.\ B, C, and E) RXTE data taken on the same day are available. Details on the observation with the longest exposure taken on the same day as an \xmm\ observation are given in table~\ref{Tab:ObsRxte}.  For the remaining two \xmm\ observations (Obs.\ A and D) the RXTE observation located closest in time is two days away. We refrain from including these observations in our study.

\subsubsection{Timing analysis}
The variability study of the RXTE observations is based on data from the Proportional Counter Array (PCA). We computed power density spectra (PDS) for each observation following the procedure outlined in \citet{2006MNRAS.367.1113B}. PDS production has been limited to the PCA channel band 0 -- 35 (2 -- 15 keV) and used 16 second long stretches of Event mode data. As for the \xmm\ data, we subtracted the contribution due to Poissonian noise  \citep{1995ApJ...449..930Z}, normalised the PDS according to \citet{1983ApJ...272..256L} and converted to square fractional rms \citep{1990A&A...227L..33B}.

\begin{minipage}[t]{\linewidth}
\resizebox{\hsize}{!}{\includegraphics[clip,angle=-90]{plot_PDS_11_35_B.ps}}\\
\resizebox{\hsize}{!}{\includegraphics[clip,angle=-90]{plot_PDS_11_35_C.ps}}\\
\resizebox{\hsize}{!}{\includegraphics[clip,angle=-90]{plot_PDS_11_35_E.ps}}

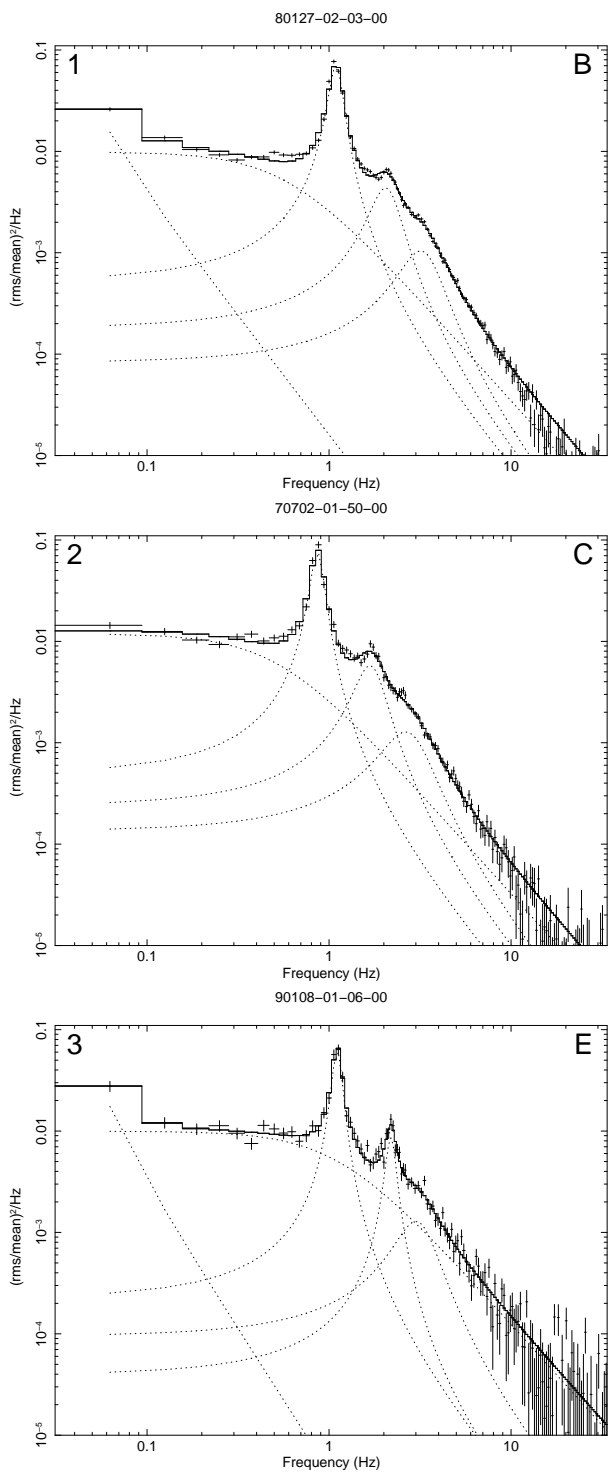
\captionof{figure}{Power density spectra of the three RXTE observations in the 4.9 -- 14.8 keV band. The best fit model is indicated by a solid line and the individual components are given as dashed lines. The letter in the upper right corner indicates the corresponding \xmm\ observation.}
\label{Fig:PDS_pca}
\end{minipage}

\subsubsection{Spectral investigation}
We used the PCA Standard 2 mode (STD2), which covers the 2--60 keV energy range with 129 channels for the spectral analysis. The standard RXTE software within \textsc{heasoft} V.~6.13 was used to extract background and dead-time corrected energy spectra for each observation, following \citet{2012MNRAS.422..679S}. Solely Proportional Counter Unit 2 from the PCA was used since only this unit was on during all the observations. To account for residual uncertainties in the instrument calibration a systematic error of 0.6 per cent was added to the PCA spectra\footnote{A detailed discussion on PCA calibration issues can be found at: http://www.universe.nasa.gov/xrays/programs/rxte/pca/doc/rmf/pcarmf-11.7/}. 

\section[]{Results}
\label{Sec:res}
\begin{figure}
\resizebox{\hsize}{!}{\includegraphics[clip,angle=0]{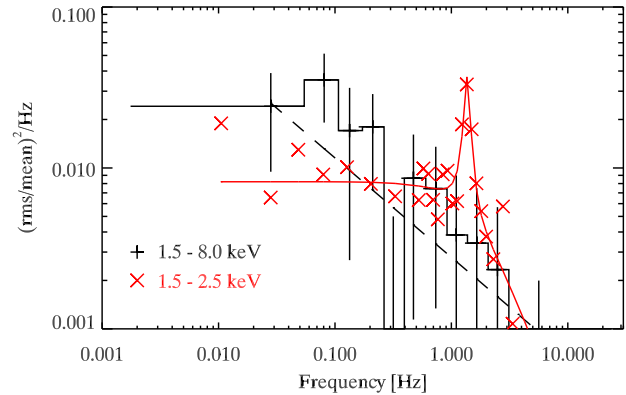}}
\caption{Averaged power density spectrum of all five observations in the 1.5 -- 2.5 keV band. The overall shape of the PDS in this energy range can be well described by a power law (dashed line). For comparison the averaged PDS of the 1.5 -- 8 keV band (red `X') are given. It is evident that a QPO is present in the 1.5 Ð 8 keV band (rms: $11.5^{+1.4}_{-1.2}$\%), while in the soft band only an upper limit of the QPO amplitude of 7.2\% rms can be derived.}
\label{Fig:PDSlow}
\end{figure}

\subsection{Power density spectra above 2.5 keV}
We fitted the PDS using a model consisting of a zero centered Lorentzian for the BLN. In the \xmm\ 4.5 -- 8 keV band QPOs are clearly visible in each observation (see Fig.~\ref{Fig:PDS}, and table~\ref{Tab:Res}), which have been fitted with a Lorentzian centered on the QPO frequency. The centroid frequencies lie in a range of 0.6 to 1.3 Hz. The fit statistics with this model is of $\chi^2/\nu = $ 28/26, 30/25, 16/21, 17/27, 23/25 for Obs.\ A--E, respectively. In Obs.\ A and B there is an excess at low frequencies. Fitting it with a power-law yields a non-significant component and a change in the fractional rms of $\sim$ 0.3~\%  in Obs.\ B.\@ Adding a power law component in Obs.\ A affects the fit of the whole spectrum and leads to a decrease of the overall variability by $\sim$ 3.0~\%.

The feature observed in the PDS of the 4.5 -- 8 keV band (BLN and QPOs) are also detected in the 2.5 -- 3.5 keV and 3.5 -- 4.5 keV bands, and the model used for the highest energy band gives also decent fits of the PDS in these bands. In these energy ranges no additional power-law component is needed in Obs.\ A and B. The parameters obtained for the BLN and QPO in all three energy bands above 2.5 keV are given in table~\ref{Tab:Res}. 

In Fig.~\ref{Fig:rmsspec} rms spectra of all five \xmm\ observations and for comparison the rms spectra of the RXTE observations are shown. The rms spectra of the main QPO obtained form RXTE data show a monotonic increase with energy and a flattening towards higher energies \citep[which has been observed in pervious studies; \eg\ ][]{2004ApJ...615..416R}. In Obs.\ A, C, and E the rms variability is highest in the 3.5 -- 4.5 keV band. While the rms values obtained from \xmm\ and RXTE in Obs.\ B are consistent within errors, the \xmm\ values in Obs.\ E are systematically higher than those obtained from RXTE data. The discrepancy of the \xmm\ and RXTE rms values in the 3.5 -- 4.5 keV band in Obs.\ C is most likely related to the poor determinability of the contribution of the BLN component in this energy band. In Obs.\ D we observe a monotonic increase in the rms variability of the main QPO peak with energy. We also noticed that in the  2.5 -- 3.5 keV band the QPO peak is broader than in the 4.5 -- 8 keV band. 

Inspired by the results obtained for the PDS below 2.5 keV (see Sect.~\ref{SubSec:PDSlow}), we tried to fit the PDS in the 2.5 -- 3.5 keV band using a model, where we substitute the zero centered Lorentzian for the BLN with a power-law component. With this substitution the Lorentzian competent to fit the upper harmonic is no longer needed. The rms variability of the upper harmonic in Obs.\ D and E is lowest in the 3.5 -- 4.5 keV band, and a similar behaviour is observed in Obs.\ B. For all observations the centroid QPO frequency stays constant within errors in all bands.

In addition, we investigated the PDS of the three RXTE observations in the 4.9 to 14.8 keV range unsung the same frequency range as for the \xmm\ observations ($1.8\times10^{-3}$ -- 34 Hz). The PDS are shown in Fig.~\ref{Fig:PDS_pca}, and their overall shape agrees with the one found from \xmm\ data. In the RXTE data, an upper harmonic is present in all three observations (parameters are given in table \ref{Tab:Res}), although it is less prominent in the first two observations than in the observation corresponding to Obs.~E, where the upper harmonic is seen with \xmm. In the third RXTE observation an additional power-law component was needed to obtain an acceptable fit at the lowest frequencies. All three RXTE observations require the addition of a second upper harmonic to obtain acceptable fits.

\begin{table*}
\caption{Variability parameters at energies above 2.5 keV}
\begin{center}
\begin{tabular}{rrrrrrrrr}
\hline\noalign{\smallskip}
\multicolumn{1}{c}{No.} & \multicolumn{1}{c}{rms$^{*}$} & \multicolumn{1}{c}{$\nu_{\mr{b}}^{\dagger}$}  &  \multicolumn{1}{c}{$\nu_{\mr{QPO}}^{\ddagger}$}  &  \multicolumn{1}{c}{QPO rms} &  \multicolumn{1}{c}{QPO HWHM$^{\star}$}  &  \multicolumn{1}{c}{$\nu_{\mr{QPO}}^{\diamondsuit}$}  &  \multicolumn{1}{c}{QPO rms} &  \multicolumn{1}{c}{QPO HWHM$^{\heartsuit}$}\\
\multicolumn{1}{c}{} & \multicolumn{1}{c}{\%} & \multicolumn{1}{c}{Hz}  &  \multicolumn{1}{c}{Hz}  &  \multicolumn{1}{c}{\%} &  \multicolumn{1}{c}{Hz}  &  \multicolumn{1}{c}{Hz}  &  \multicolumn{1}{c}{\%} &  \multicolumn{1}{c}{Hz}\\
\hline\noalign{\smallskip}
\multicolumn{9}{c}{4.5 -- 8 keV}\\
\hline\noalign{\smallskip}
A & $20.8^{+6.0}_{-6.6}$& $1.31^{+0.79}_{-0.57}$& $0.90\pm0.01$& $12.2^{+1.2}_{-1.5}$& $0.12\pm0.01$& -- & -- & -- \\
\noalign{\smallskip} 
B & $25.9^{+3.3}_{-3.6}$& $2.16^{+0.79}_{-0.58}$& $1.15\pm0.01$& $14.7\pm1.1$& $0.08\pm0.02$& -- & -- & -- \\
\noalign{\smallskip} 
C & $11.7^{+3.6}_{-4.2}$& $0.10^{+0.14}_{-0.06}$& $0.72^{+0.02}_{-0.03}$ & $11.4^{+2.2}_{-2.4}$ & $0.05\pm0.04$& -- & -- &-- \\
\noalign{\smallskip} 
D & $16.0^{+3.9}_{-5.7}$& $1.05^{+0.90}_{-0.63}$& $0.61\pm0.01$& $10.0\pm0.9$& $0.05^{+0.02}_{-0.01}$&$1.15^{+0.04}_{-0.03}$ & $8.8^{+1.8}_{-1.6}$& $0.10^{+0.06}_{-0.05}$ \\
\noalign{\smallskip} 
E & $>13.4$& $0.51^{+0.69}_{-0.21}$& $1.38\pm0.02$& $12.9^{+1.0}_{-1.6}$& $0.10\pm0.03$&$2.81^{+0.09}_{-0.10}$ & $5.9^{+2.1}_{-2.3}$& $0.12^{+0.30}_{-0.12}$  \\
\hline\noalign{\smallskip} 
\multicolumn{9}{c}{3.5 -- 4.5 keV}\\
\hline\noalign{\smallskip}
A & $13.2^{+2.5}_{-2.4}$& $0.10^{+0.07}_{-0.04}$& $0.92\pm0.06$& $16.2^{+2.2}_{-2.0}$& $0.24^{+0.13}_{-0.08}$& -- & -- & -- \\
\noalign{\smallskip} 
B & $19.6^{+3.0}_{-3.1}$& $0.30^{+0.15}_{-0.10}$& $1.14^{+0.03}_{-0.07}$& $8.7^{+2.0}_{-2.4}$& $<0.1$& -- & -- & -- \\
\noalign{\smallskip} 
C & -- & -- & $0.77^{+0.03}_{-0.04}$ & $21.2^{+2.6}_{-2.9}$ & $0.08\pm0.03$& -- & -- &-- \\
\noalign{\smallskip} 
D & $> 9.3$& $> 0.08$& $0.64\pm0.03$& $8.9^{+1.9}_{-3.6}$& $0.07^{+0.05}_{-0.06}$&$1.18^{+0.01}_{-0.12}$ & $6.1^{+1.3}_{-1.6}$& $<0.21$ \\
\noalign{\smallskip} 
E & $12.7^{+5.8}_{-7.6}$& $0.48^{+0.75}_{-0.45}$& $1.32^{+0.05}_{-0.06}$& $14.1^{+2.9}_{-2.3}$& $0.17^{+0.13}_{-0.07}$&$2.74^{+0.04}_{-0.22}$ & $4.7^{+1.5}_{-2.3}$& $<0.08$  \\
\hline\noalign{\smallskip} 
\multicolumn{9}{c}{2.5 -- 3.5 keV}\\
\hline\noalign{\smallskip}
A & $12.3^{+2.5}_{-2.6}$& $0.14^{+0.10}_{-0.06}$& $0.91^{+0.07}_{-0.05}$& $11.7^{+1.9}_{-2.0}$& $0.14^{+0.08}_{-0.06}$& -- & -- & -- \\
\noalign{\smallskip} 
B & $12.6^{+2.3}_{-2.5}$& $0.09^{+0.07}_{-0.04}$& $1.12^{+0.05}_{-0.06}$& $10.9^{+1.8}_{-2.1}$& $0.10^{+0.07}_{-0.06}$& -- & -- & -- \\
\noalign{\smallskip} 
C & $18.4^{+5.9}_{-7.5}$ &$0.17^{+0.21}_{-0.13} $ & $0.73^{+0.02}_{-0.01}$ & $9.6^{+2.8}_{-2.7}$ & $<0.05$& -- & -- &-- \\
\noalign{\smallskip} 
D & $11.2^{+3.9}_{-4.3}$& $0.14^{+0.22}_{-0.09} $& $0.64^{+0.04}_{-0.05}$& $8.3^{+2.9}_{-4.3}$& $<0.18$ &$1.46^{+0.40}_{-0.46}$ & $12.8^{+5.7}_{-4.3}$& $0.55^{+2.23}_{-0.41}$ \\
\noalign{\smallskip} 
E & $15.6^{+5.8}_{-4.9}$& $0.59^{+0.65}_{-0.32}$& $1.31^{+0.05}_{-0.08}$& $11.6^{+3.0}_{-3.2}$& $0.16^{+0.15}_{-0.10}$&$2.77^{+0.01}_{-0.02}$ & $7.5^{+1.3}_{-1.6}$& $<0.06$  \\
\hline\noalign{\smallskip} 
\multicolumn{9}{c}{4.9 -- 14.8 keV (RXTE/PCA)}\\
\hline\noalign{\smallskip}
1 & $13.5\pm0.2$& $0.59\pm0.03$& $1.092\pm0.002$& $14.4\pm0.1$& $0.093^{+0.002}_{-0.003}$& $2.04\pm0.01$ & $7.4\pm0.2$ & $0.42^{+0.03}_{-0.02}$ \\
\noalign{\smallskip} 
2 & $13.8^{+0.5}_{-0.4}$& $0.51^{+0.05}_{-0.03}$& $0.865^{+0.003}_{-0.002}$& $13.0^{+0.2}_{-0.3}$& $0.067^{+0.006}_{-0.004}$& $1.67\pm0.02$ & $7.7^{+0.4}_{-0.5}$ & $0.35\pm0.05$ \\
\noalign{\smallskip} 
3 &  $18.7^{+0.7}_{-0.6}$& $1.12^{+0.09}_{-0.07}$& $1.109\pm0.005$& $11.6\pm0.3$& $0.06\pm0.01$& $2.19\pm0.01$ & $6.3\pm0.3$ & $0.15\pm0.03$ \\
\hline\noalign{\smallskip}
\end{tabular} 
\end{center}
Notes: $^{*}$: variability of the band limited noise in the $1.8\times10^{-3}$ -- 34 Hz range\\
$^{\dagger}$: break frequency of the band limited noise\\
$^{\ddagger}$: centroid frequency of the (fundamental) QPO\\
$^{\star}$: Half width at half maximum of the (fundamental) QPO\\
$^{\diamondsuit}$: centroid frequency of the (harmonic) QPO\\
$^{\heartsuit}$: Half width at half maximum of the (harmonic) QPO\\
\label{Tab:Res}
\end{table*}

\subsection{Power density spectra below 2.5 keV}
\label{SubSec:PDSlow}
For the lowest energy band (1.5 -- 2.5 keV), we had to average the PDS of all five observations to obtain decent fit statistics. Assuming a power-law noise, we obtained a decent fit, without statistical need of a zero centered Lorentzian component, with $\chi^2/\nu = $ 16/14 and a power-law index of $0.7_{-0.1}^{+0.2}$. Futhermore, the QPO, seen clearly at energies above 2.5 keV, seems not to be present in this energy band. The averaged PDS of the 1.5 -- 2.5 keV band is shown in Fig.~\ref{Fig:PDSlow} together with the averaged PDS of the 1.5 -- 8 keV band. Adding a Lorentzian component with the centroid frequency (1.37 Hz) and width fixed at the values obtained from a fit to the averaged PDS in the 1.5 -- 8 keV band, we obtain a 1 $\sigma$ upper limit for the QPO amplitude of 7.2 \% rms. This value lies a little bit below the QPO amplitude value at soft energies obtained from \citet{2004ApJ...615..416R} for a similar centroid frequency; $\sim$ 8 -- 9 \% rms. However, the upper limit we obtained from the \xmm\ data is not very stringent. We also tried to fit the PDS with a zero centered Lorentzian instead of the power-law, which results in a $\chi^2/\nu$ of 11/14. We find a break frequency of the Lorentzian of 0.45$^{+0.65}_{-0.21}$ Hz, which is clearly below the break frequency of the Lorentzian in the 1.5 -- 8 keV band (3.35$^{+0.52}_{-0.49}$ Hz). 

\begin{figure*}
\resizebox{\hsize}{!}{\includegraphics[clip,angle=-0]{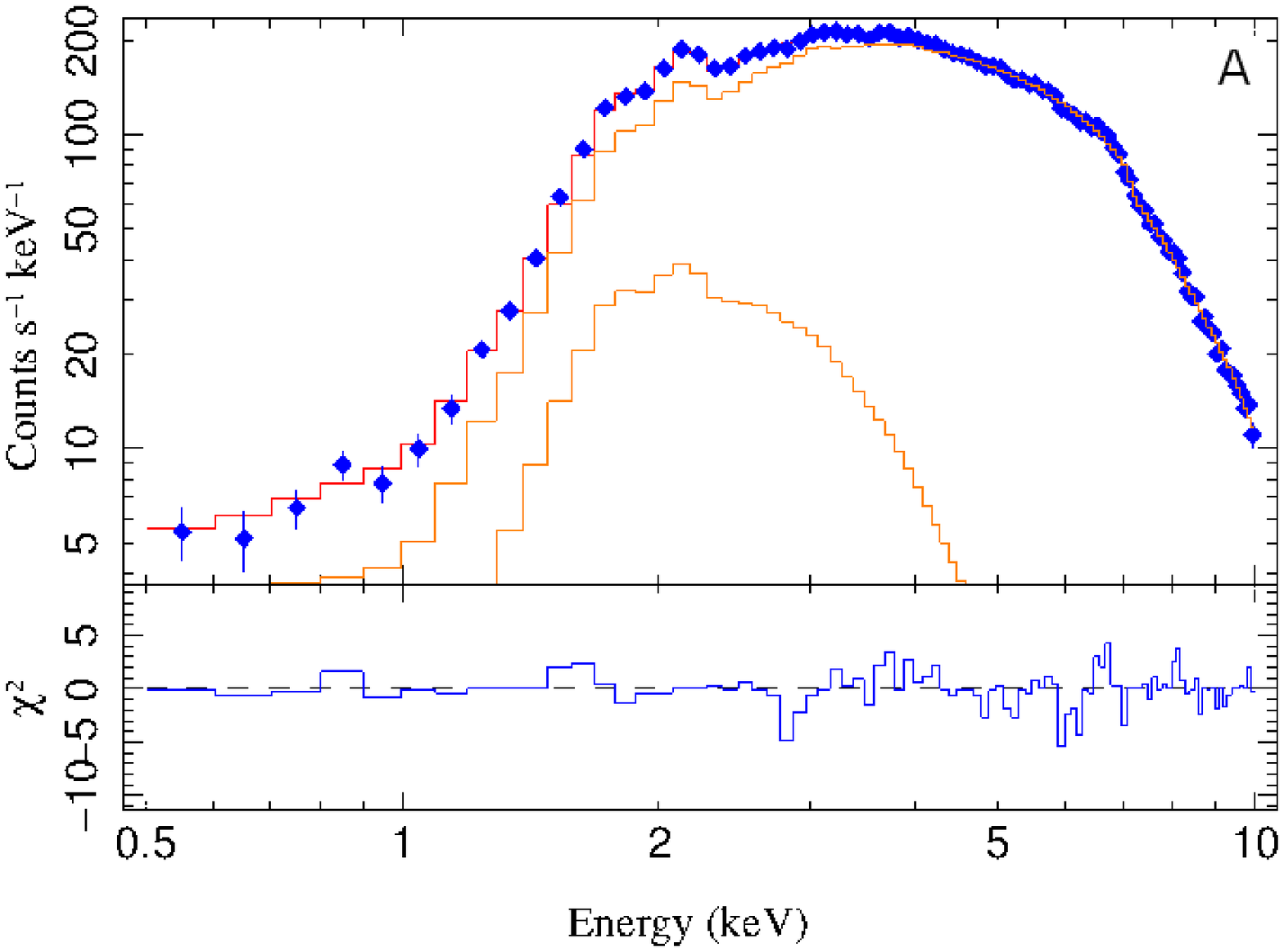}\hskip0.2cm\includegraphics[clip,angle=-0]{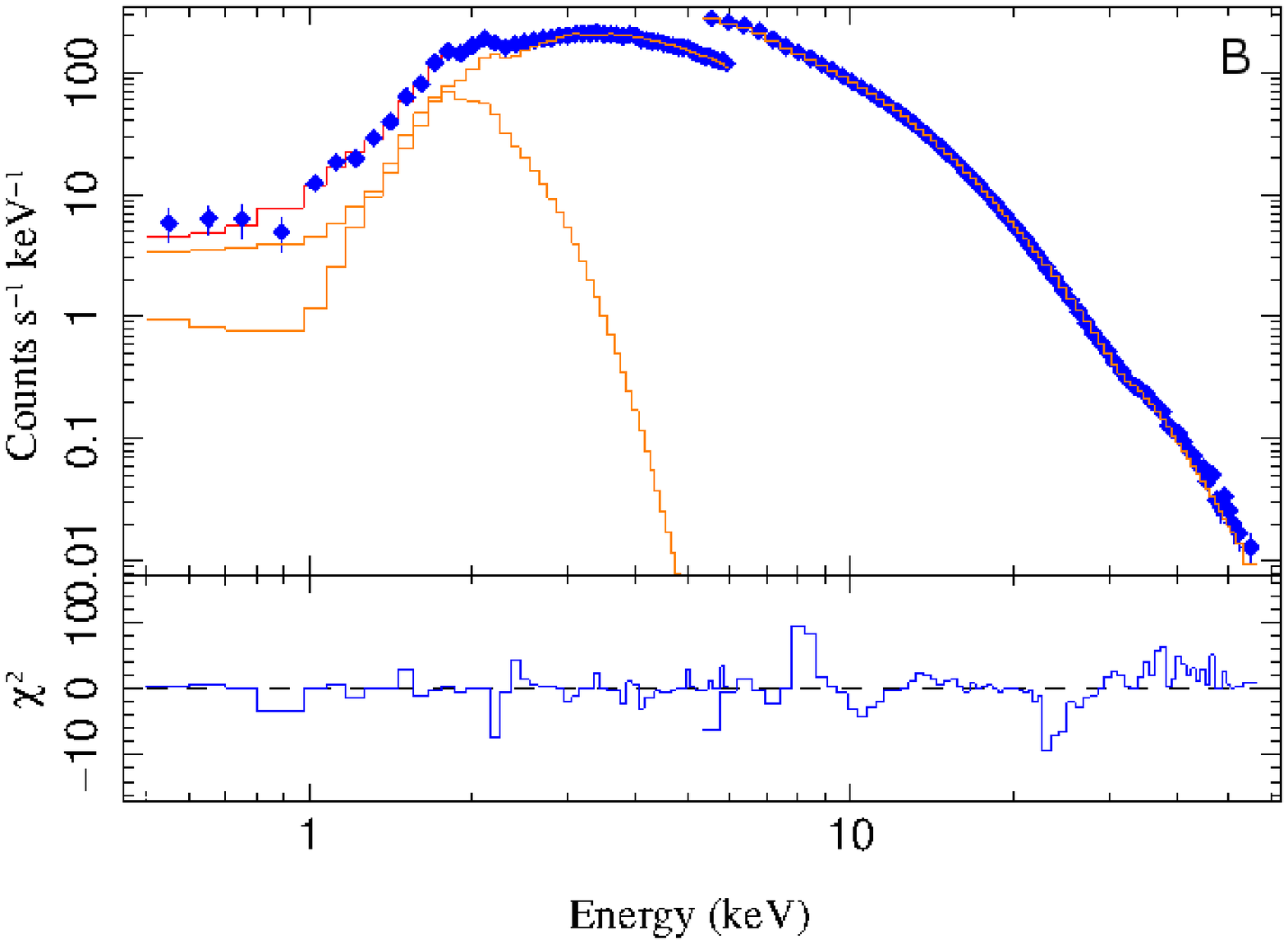}}
\resizebox{\hsize}{!}{\includegraphics[clip,angle=-0]{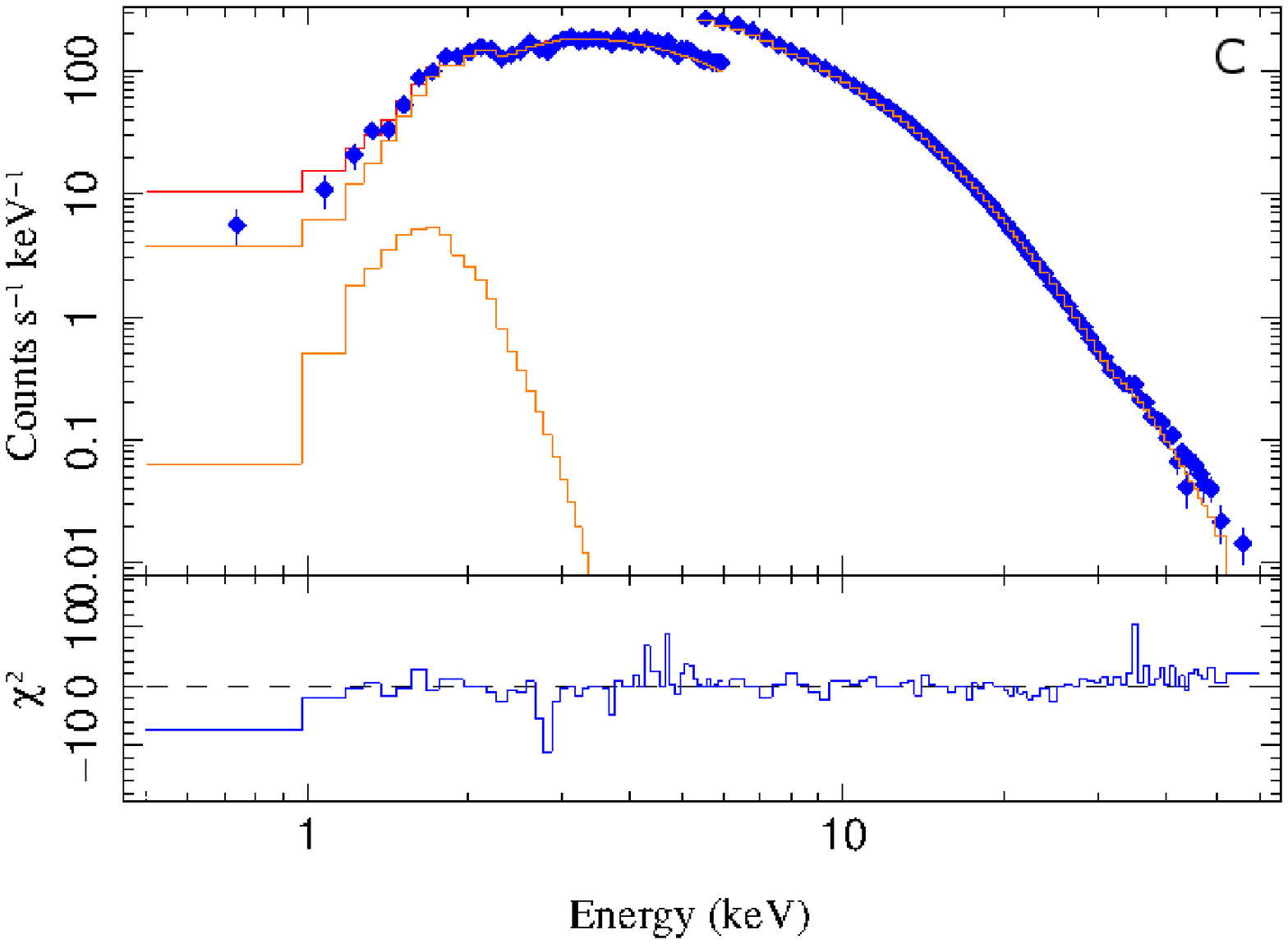}\hskip17.9cm}\\
\resizebox{\hsize}{!}{\includegraphics[clip,angle=-0]{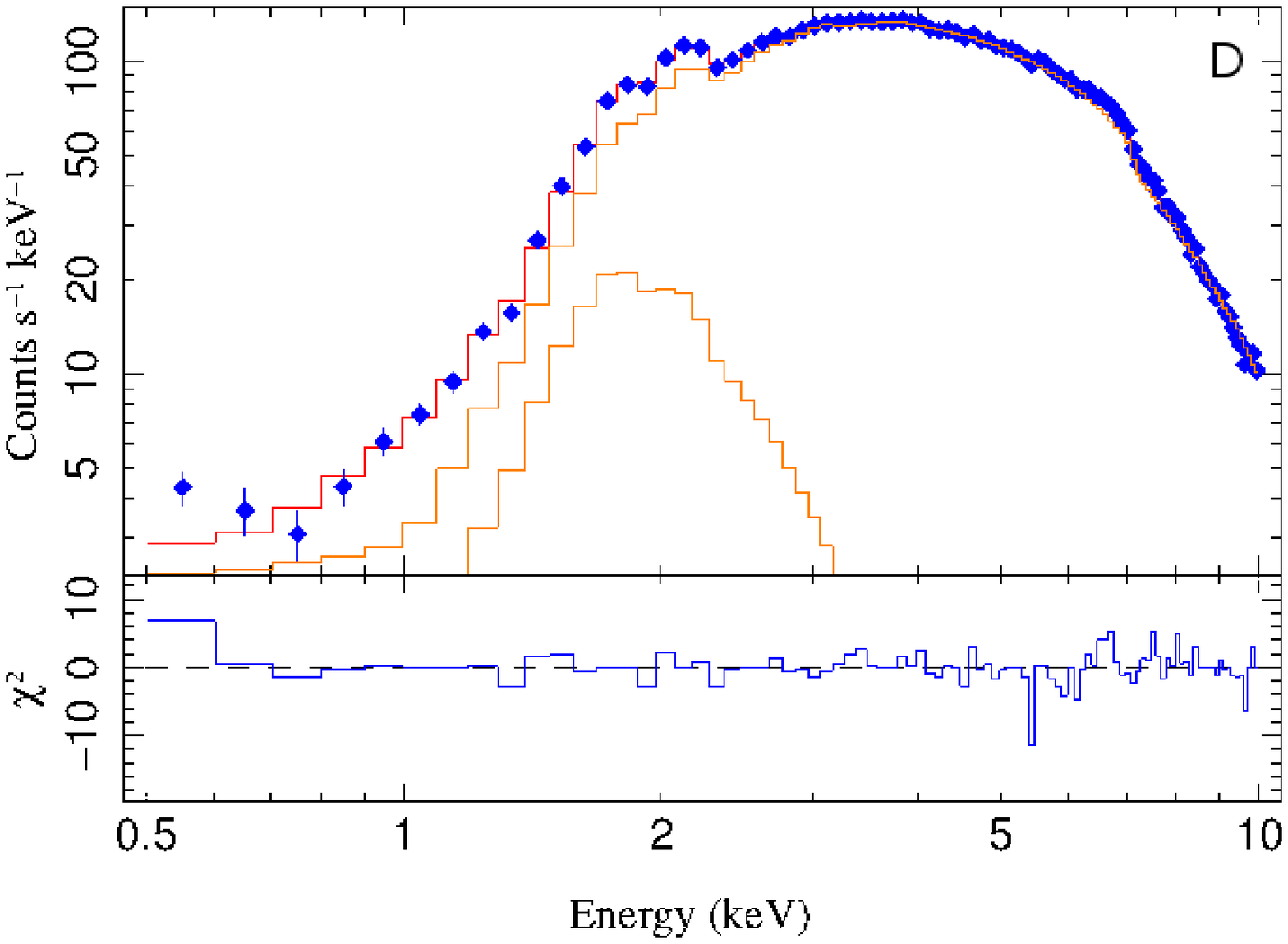}\hskip0.2cm\includegraphics[clip,angle=-0]{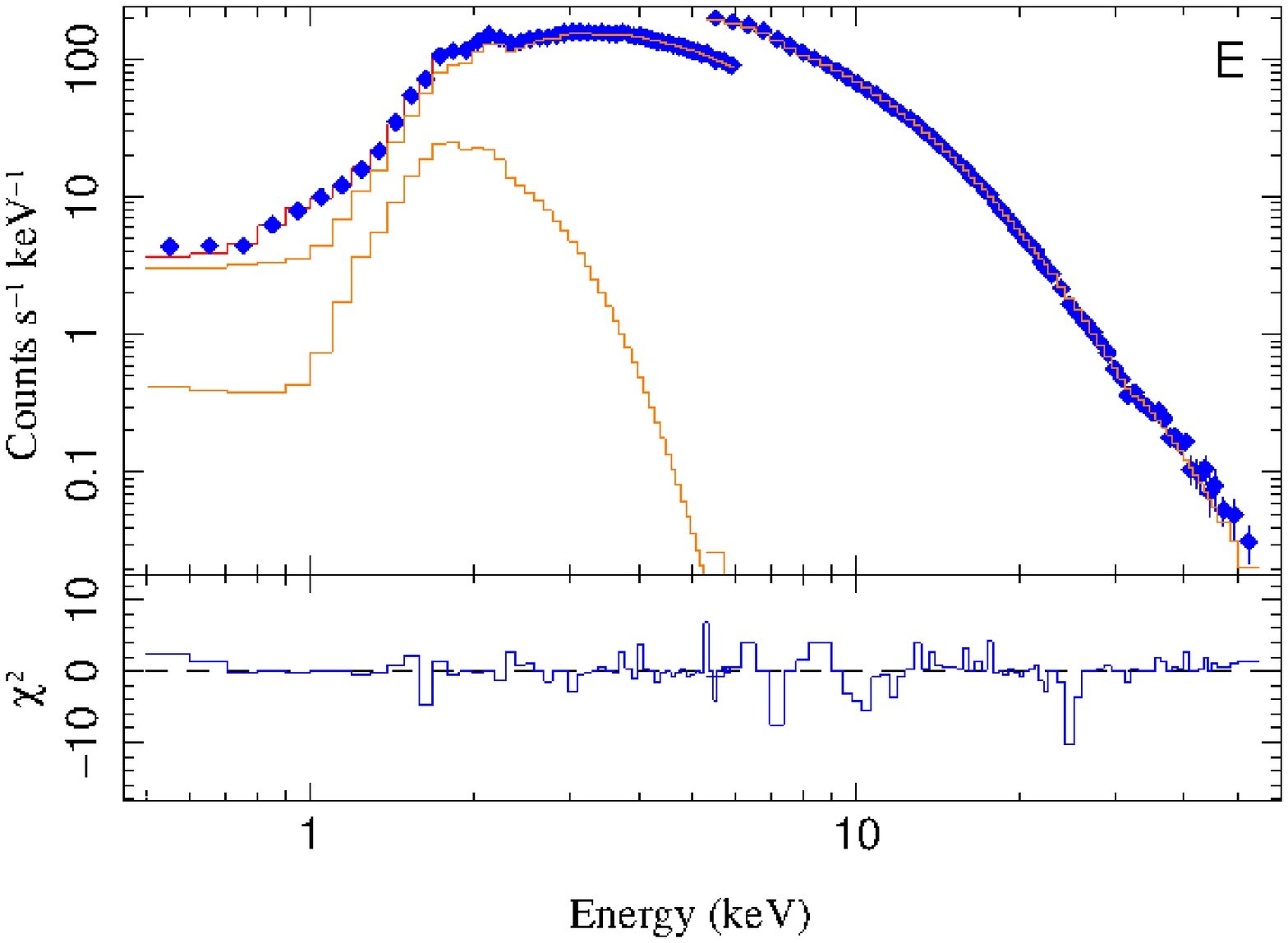}}
\caption{Energy spectra and residuals of all five \xmm\ observations. For Obs.\ B, C, and E the simultaneous RXTE/PCA data are included in the fit. The best fit model as well as the contribution of the disc blackbody emission and of the power law with reflection component are indicated by solid lines.}
\label{Fig:Spec}
\end{figure*}

\begin{table}
\caption{Selected spectral parameters}
\begin{center}
\begin{tabular}{rrlrr}
\hline\noalign{\smallskip}
 \multicolumn{1}{c}{No.} & \multicolumn{1}{c}{$\Gamma$} & \multicolumn{1}{c}{T$_{\mr{in}}$}  &   \multicolumn{1}{c}{R$_{\mr{in}}\times D_{10}$}  \\
  &  & \multicolumn{1}{c}{keV}  &  \multicolumn{1}{c}{km}  \\
 \hline\noalign{\smallskip}
A &  1.37$^{+0.03}_{-0.02}$ & 0.56$^{+0.05}_{-0.06}$ &  82$_{-12}^{+21}$\\
\noalign{\smallskip} 
B & 1.85$^{+0.02}_{-0.03}$ & 0.23$\pm$0.01 & 3179$_{-560}^{+555}$\\
\noalign{\smallskip} 
C & 1.62$\pm$0.03 & 0.18$^{+0.06}_{-0.05}$& 1886$_{-1515}^{+0}$\\
\noalign{\smallskip} 
D & 1.32$^{+0.01}_{-0.02}$ & 0.32 $^{+0.05}_{-0.02}$&337$_{-133}^{+86}$ \\
\noalign{\smallskip} 
E & 1.64$\pm$0.02 & 0.33$^{+0.02}_{-0.01}$&313$_{-44}^{+51}$ \\
\hline\noalign{\smallskip} 
\end{tabular} 
\end{center}
\label{Tab:SpecPar}
\end{table}

\subsection {Spectral results}
Furthermore, the presence of a power-law component in the PDS below 2.5 keV, which is commonly observed in the HSS where the energy spectrum is dominated by emission form the accretion disc, suggests that a disc component should be present in the energy spectra, as inferred from the study of MAXI J1659-152 \citep{2013ApJ...770..135Y}. This is in contrast to the results presented in \citet{2006A&A...448..677M}, where only a reflected power-law component (but no direct disc emission) was need to obtain acceptable fits, using solely \xmm\ data. Hence, we fitted combined \xmm/EPIC-pn+RXTE/PCA spectra within \textsc{isis} V.~1.6.2 \citep{2000ASPC..216..591H} in the 0.5 -- 10 keV and 5 -- 60 keV range. We fitted the spectra with the model used in \citet{2006A&A...448..677M}, consisting of  a power law with reflection component (\textsc{refsch} in \textsc{Xspec}) modified by cold absorption, several emission features, and an additional component to model the 1 keV excess. Using an unmodified \xmm\ spectrum plus an RXTE/PCA spectrum or an \textsc{epfast}-corrected \xmm\ spectrum plus an RXTE/PCA spectrum, the features present in the data do not allow us to obtain a reduced $\chi^2$ below 2 (see Sect.~\ref{SubSec:XMMSpec}). To get formally acceptable fits we applied \textsc{epfast} to the EPIC-pn spectrum and ignored energies above 6 keV in this spectrum, while the RXTE/PCA spectrum was used in the 5 -- 60 keV range. For Obs.\ A and D where no simultaneous RXTE data are available \xmm\ data are used up to 10 keV. We grouped the \xmm\ data to contain at least 20 channels per bin. Both \xmm\ and RXTE data are grouped to have a signal-to-noise ratio larger than three per bin.  

The energy spectra and residuals of all five observations with their best fit model are shown in Fig.~\ref{Fig:Spec}.The $\chi^2/\nu$ are 98/95, 202/130, 172/123, 137/95, and 150/126 for Obs.\ A, B, C, D, and E, respectively. The fold energy obtained form fits of combined \xmm\ and RXTE data are outside the energy range covered by \xmm\ (in the range of 11 -- 12 keV for Obs.\ B and C and $\sim$ 14 keV for Obs.\ E). That is why we fixed the fold energy in Obs.\ A and D at 11 and 14 keV, respectively. The obtained photon index, inner disc temperature and inner disc radius are given in tabel~\ref{Tab:SpecPar}.
The disc component always peaks around 1.5 -- 2.5 keV. In most observations the flux from the disc component in the 1.5 -- 2.5 keV band makes up about 20 -- 30 \% of the total flux, while in the 2.5 -- 3.5 keV band the contribution of the disc component reduces to a few per cent. This finding puts additional support on the presence of a power-law component in the PDS below 2.5 keV, as we found a power-law noise in MAXI J1659-152 at a disc fraction exceeding $\sim$ 30~\% in the 0.3 -- 2 keV band \citep{2013ApJ...770..135Y}. It is worth noting that an extension of the power-law spectral component to soft energies is not physical. Using a simple power-law spectral component in the spectral fit would actually underestimate the disc component.

We used Obs.\ E, that has the longest exposure of all three \xmm\ observations with simultaneous RXTE data, to verify that in all three cases -- unmodified \xmm\ spectrum plus RXTE/PCA spectrum, \textsc{epfast}-corrected \xmm\ spectrum plus RXTE/PCA spectrum, and \textsc{epfast}-corrected \xmm\ spectrum below 6 keV plus RXTE/PCA spectrum above 5 keV -- the addition of a multicolor disc blackbody component leads to a significant improvement of the fit.

\section[]{Discussion \& Conclusion}
\label{Sec:dis}
We studied studied archival \xmm\ data of \grs\ during its $\chi$ variability class obtained in 2003 and 2004. The focus of our study was put on an investigation of the power spectral shape in different X-ray energy bands, in the light of our work done with \textit{Swift} on MAXI J1659-152 \citep{2013ApJ...770..135Y}. We found that while the PDS at energies above 2.5 keV is dominated by BLN plus QPO, corresponding to the power spectral shape seen in the hard or intermediate state, the PDS in the energy range between 1.5 and 2.5 keV shows power-law noise, which corresponds to the power spectral shape usually seen in the soft state.
A similar existence of two distinct power spectral states, BLN plus QPO above 2 keV and PLN below 2 keV, has been found for MAXI J1659--152 in its hard intermediate state \citep{2013ApJ...770..135Y}. Our result that the PDS of the $\chi$ class of \grs\ shows a similar energy dependence as the hard intermediate state in MAXI J1659--152 fits well into the known connection of the $\chi$ variability class in \grs\ with the intermediate state (preferentially close to the hard state) in other black hole X-ray binaries \citep{2003A&A...412..229R,2013ApJ...778..136P}.

In the study of MAXI J1659-152, \citet{2013ApJ...770..135Y} have found that the power-law noise in the soft band seems to have a cut-off at or below the QPO and BLN break frequency. The \xmm\ data of \grs\ do not allow us to determine if there is such a cut-off. Studying RXTE PDS in individual energy bands, only those of Obs.~3 (the one corresponding to \xmm\ Obs.~E) show at energies below 4.5 keV a clear deviation from a BLN plus QPO shape at frequencies below $\sim$0.2 Hz that can be described by a power-law component. The frequency at which the deviation from the BLN plus QPO shape occurs decreases with increasing energy. However, this finding does not allow to draw conclusions on a possible cut-off in the 1.5 -- 2.5 keV band, as the band covered by RXTE is too broad (up to 4.5 keV; and data with a higher energy resolution are not available) and the presence of an additional power-law noise is only found in one out of three RXTE observations.

In summary, the observations of \grs\ show similar energy-dependent power spectral states as in MAXI J1659-152, which means that two different power spectral shapes are coexisting in the hard and soft band simultaneously. In the soft band, which is contributed by emission from the thermal disc component, not only the variability amplitude is lower, as has been known before, but the power spectral shape is of a power-law shape. Such an energy dependence reveals a geometry in which the photons in the soft energy band and in the hard energy band come from different locations in the system, \ie\ the cold optically thick accretion disc and the region of Comptonization of hot electrons (being it either an optically thin hot corona \citep[see \eg][]{1997ApJ...489..865E} or a jet flow \citep[see \eg][]{2005ApJ...635.1203M}), respectively. The inner radius at which the cold disc component ends would be determined by future accurate measurements of the power-law noise in the soft band. Notice that the radius at which the cold disc ends may be not the radius at which the cold disc terminates \citep[as assumed in the truncation disc model; \eg][]{2001A&A...373..251D,2006csxs.book..157M,2007A&ARv..15....1D}, since a hot flow or corona would cover the innermost cold disc so the radius one determines from the cut-off frequency of the power-law noise would correspond to the radius to which the hot flow or corona extends \citep[while the disc can reach down to the ISCO as assumed \eg\ in][]{1999ApJ...510L.123B,2002MNRAS.332..165M,2006ApJ...653..525M,2013ApJ...763...48R}. 

In conclusion, the energy dependence of the power spectral state found supports the idea that the observed power spectral state depends on which spectral component (and thus the geometrical location -- disc or corona) we are looking at, and that a multi-wavelength picture of power spectra in black hole X-ray binaries is needed. The important consequence of such an energy-dependent picture of black hole power spectral state is that power spectral analysis in the soft X-ray energy band is more sensitive to the emergence of the disc component in the hard or intermediate state than energy spectral analysis, which in many circumstances is model dependent.  

\section*{Acknowledgments}
We would like to thank Tomaso Belloni, Masaru Matsuoka, Phil Kaaret, Mike Nowak, and Deepto Chakrabarty for comments and useful discussions. This work was supported by the National Natural Science Foundation of China under grant No. 11073043, 11333005, and 11350110498, by Strategic Priority Research Program "The Emergence of Cosmological Structures" under Grant No. XDB09000000 and the XTP project under Grant No. XDA04060604, by the Shanghai Astronomical Observatory Key Project and by the Chinese Academy of Sciences Fellowship for Young International Scientists Grant. 

\bibliographystyle{mn2e}
\bibliography{/Users/holger/work/papers/my2010,/Users/holger/work/papers/my2013}

\appendix

\bsp

\label{lastpage}

\end{document}